\documentclass[11pt]{article}
\usepackage{float,epsfig,epstopdf}
\usepackage{graphicx}
\usepackage{amsmath}
\usepackage{amssymb,multirow}
\usepackage{bm}
\usepackage{tabularx,ragged2e,booktabs,caption}
\usepackage{setspace}
\usepackage{multirow}

\begin{document}
\begin{center}
{\LARGE\bf  A new robust class of skew elliptical distributions}
\\[1ex]
by
\\[1ex]
Hok Shing Kwong and Saralees Nadarajah
\\
Department of Mathematics, University of Manchester, Manchester M13 9PL, UK
\\[1ex]
\end{center}
\vspace{1.5cm}
{\bf Abstract:}~~A new robust class of multivariate skew distributions is introduced.
Practical aspects such as parameter estimation method of the proposed class are discussed,
we show that the proposed class can be fitted under a reasonable time frame.
Our study shows that the class of distributions is capable to model multivariate skewness structure
and does not suffer from the curse of dimensionality as heavily as other distributions of similar complexity do,
such as the class of canonical skew distributions.
We also derive a nested form of the proposed class which appears to be the
most flexible class of multivariate skew distributions in literature that has a closed-form density function.
Numerical examples on two data sets, i) a data set containing daily river flow data recorded in the UK;
and ii) a data set containing biomedical variables of athletes collected by the Australian Institute of Sports (AIS), are demonstrated.
These examples further support the practicality of the proposed class on moderate dimensional data sets.


\section{Introduction}

Probability distributions are the core of parametric statistical analysis.
Throughout the past decades, there has been a growing interest in the development of flexible parametric families of distributions.
This surge in literature was partly caused by the increased needs
for distributions to capture non-normalities in data.
With the aid of flexible distributions, unrealistic assumptions such as symmetry and normality can be avoided,
hence, more robust statistical analysis can be carried out.

Leptokurtosis and asymmetry are two main sources of non-normality.
To model leptokurtosis, distributions such as Student's $t$-distribution, power exponential distribution,
and their elliptical counter parts (Kelker, 1970) (see, for example, Cambanis et al. (1981) and Fang et al. (1990) for systematic reviews)
are found to be useful, and in many cases, adequate to model leptokurtosis.
On the other hand, modelling asymmetry is often regarded as a trickier task,
probably due to the complexity of multivariate asymmetry.
As a result, the development of asymmetric distributions has attracted a great deal of attention among researchers.

The most prominent class of multivariate skew distributions is arguably the skew-symmetric family, constructed based on symmetric distributions.
To date, many different skew-symmetric distributions have been proposed and studied in detail.
Here is a brief summary of the progress made: Azzalini and Dalla-Valle (1996),
and Azzalini and Capitanio (1999) introduced the first multivariate skew-normal distribution;
Branco and Dey (2001) proposed a general class of skew-elliptical distributions constructed
from hidden truncations of elliptical distributions where the multivariate skew-normal distribution
by Azzalini and Dalla-Valle (1996) is a special case;
Arnold and Beaver (2000) introduced the skew-Cauchy distribution;
Genton and Loperfido (2005) derived the generalised skew-elliptical distributions admitting a variety of choices for skewing functions;
Sahu et al. (2003) derived another form of skew-elliptical distribution capable of capturing multivariate asymmetry;
Azzalini and Capitanio (2003) studied a general procedure to obtain skew-symmetric distributions;
Wang et al. (2004) introduced a skew-symmetric representation;
Ma and Genton (2004) studied a flexible class of skew-symmetric distributions in which
skewing functions are constructed from polynomials;
Arellano-Valle and Genton (2005) proposed the so-called fundamental skew distributions, including
all of the aforementioned skew distributions as special cases;
Arellano-Valle and Genton (2010b) introduced the multivariate extended skew-$t$ distribution that can model platykurtosis.
Other papers include Ferreira et al. (2016), Kahrari et al. (2016) and Arellano-Valle et al. (2018).

Many of the existing skew-symmetric distributions have attractive statistical properties,
such as having analytically tractable pdf or simple stochastic representations.
However, we found that their practical performance often suffers heavily from the curse of dimensionality.
For example,  the class introduced in Sahu et al. (2003)
has a  computationally intensive pdf, which quickly becomes infeasible  as the number of dimensions grows.
In view of this, we believe that the development of a new class of multivariate skew distributions
that can remain computationally feasible at higher dimensions is much needed.

In this paper, we introduce a novel class of multivariate skew distributions
that has practical advantages over the previous ones, especially in higher dimensions.
The remainder of this paper is organised as follows.
Sections \ref{section:univariate} and \ref{section:multivariate} briefly review,
respectively, univariate and multivariate cases of skew-elliptical distributions.
This helps to provide a better understanding of the development of the proposed distributions.
The new class of distributions is proposed and discussed in Section \ref{section:gms}.
Numerical examples are presented in Section \ref{section:examples}.
Finally, concluding remarks are given in Section \ref{section:conclusion}.

\section{Univariate generalised skew-elliptical distributions}
\label{section:univariate}

Our starting point is the univariate case of generalised skew-elliptical distributions (GSEs).
Generally, GSE can be constructed from perturbation of elliptical distributions and its pdf has the form
\begin{align}
\displaystyle
2f(x)g(\lambda x),
\label{eq:gse}
\end{align}
where $f$ is an elliptical pdf and $g$ is a skewing function that satisfies the conditions $g(x) = 1 - g(-x)$,
and $0 \le g(x) \le 1$.
It is not difficult to show that $2f(x)g(x)$ is a valid pdf.
For instance, it can be shown that
\begin{align*}
\displaystyle
\int_{-\infty}^{\infty}g(\lambda x)f(x;\phi)dx
&=
\displaystyle
\int_{0}^{\infty}g(\lambda x)f(x;\phi)dx + \int_{-\infty}^{0}g(\lambda x)f(x;\phi)dx
\\
&=
\displaystyle
\int_{0}^{\infty}g(\lambda x)f(x;\phi)dx + \int_{-\infty}^{0}\left(1-g(-\lambda x)\right)f(x;\phi)dx
\\
&=
\displaystyle
\int_{0}^{\infty}g(\lambda x)f(x;\phi)dx + \int_{0}^{\infty}\left(1-g(\lambda x)\right)f(x;\phi)dx
\\
& =
\displaystyle
\int_{0}^{\infty}f(x;\phi)dx
\\
& = \frac {1}{2}
\end{align*}
and therefore (\ref{eq:gse}) is a valid pdf.

Many of the commonly applied skew-distributions are special cases of GSE.
The skew-normal distribution due to Azzalini (1985, 1986) is an example.
Its pdf has the form
\begin{align}
\displaystyle
f(z; \lambda) = 2\Phi(\lambda z)\phi(z),
\hphantom{1}
z\in\mathbb{R},
\hphantom{1}
\lambda\in\mathbb{R},
\label{eq:sn}
\end{align}
where $\Phi$ and $\phi$ denote, respectively,  the cdf and pdf of the standard normal distribution.

The main advantage of the skew-normal distribution is that it has a simple stochastic representation.
For example, if $\left(Y_1, Y_2\right) \sim N \left({\bf 0}, {\bf I} \right)$, (\ref{eq:sn}) can be shown as the pdf of $Z$ such that
\begin{align*}
\displaystyle
Z = \frac {1}{\sqrt{1+\lambda^2}}Y_1 + \frac {\lambda}{\sqrt{1+\lambda^2}} \left | Y_2 \right |.
\end{align*}
Moreover, if $\left(Y_1, Y_2\right) \sim t_2 \left( {\bf 0}, {\bf I}, \nu \right)$ instead, it can be shown that the pdf of $Z$ is now
\begin{align}
\displaystyle
2T_1 \left(\lambda z;\nu+1\right) t_1(z;\nu),
\label{eq:st}
\end{align}
where $T_p$ and $t_p$ denote, respectively,  the joint cdf and the joint pdf of a $p$-dimensional elliptical Student's $t$-distribution.

Skew distributions with this type of stochastic representations are sometimes referred to
as canonical skew distributions.
Statistical properties of canonical skew distributions have been studied intensively
due to this attractive stochastic representation.
However, from a practical perspective, it is not always beneficial to apply canonical skew distributions in statistical analysis.
For example, if the objective of the analysis is to approximate the distribution of some data sets,
it is reasonable to use distributions that do not have simple stochastic representations,
especially when closed-form expression of maximum likelihood estimates (MLEs) generally do not exist for skew-elliptical distributions,
and parameter estimations mostly rely on numerical methods.
That being said, if the normal cdf in (\ref{eq:sn}) or Student's $t$ cdf in (\ref{eq:st}) are replaced
by some closed-form skewing functions, we can expect the parameter estimations to be more efficient.
To illustrate the practical differences between skewing functions, a data set containing daily river
flow recorded at different stations in the UK from 01/01/2000 to 31/12/2019
\footnote{obtained from the UK environment agency website; https://environment.data.gov.uk/hydrology/explore}
were fitted to skew-$t$ distributions with different skewing functions.
Some details of the data sets are shown in Table \ref{tab:river}, and the skewing functions considered are as follows:
\begin{itemize}

\item
cdf of Student's $t$: $\frac {1}{2}+x\Gamma\left(\frac {\nu+1}{2}\right) \frac {{}_2F_1 \left(\frac {1}{2},\frac {\nu+1}{2};\frac {3}{2};-\frac {x^2}{\nu} \right)}{\sqrt{\pi\nu}\Gamma \left(\nu/2 \right)}$;

\item
error: $\frac {1}{2}\left[1+\text{erf} \left( \frac {x}{\sqrt 2} \right) \right]$;

\item
hyperbolic secant: $\frac {2}{\pi} \text{arctan} \left[ {\text{exp}(x\pi/2)} \right]$;

\item
logistic: $\frac {1}{1+e^{-x}}$;

\item
arctan: $1/2+\frac {\text{arctan}(x)}{\pi}$;

\item
reciprocal square root: $\frac {x}{2\sqrt{1+x^2}}+\frac {1}{2}$.

\end{itemize}

\begin{singlespace}
\begin{table}[H]
\centering
\begin{tabular}{cccc}
River & Station name            & River name       & WISKI ID
\\
\hline
1 &London-Road             & River Medlock    & 690713
\\
2 &South-Willesborough     & East Stour River & 654210001
\\
3&Collyhurst-Weir         & River Irk        & 690611
\\
4 & Alfoldean               & River Arun       & 254250008
\\
5 &Denham-Lodge            & River Misbourne  & 2879TH
\\
6 &Kirkby                  & River Alt        & 694744
\\
7 &Morwick                 & River Coquet     & 022001
\\
8 &Ower                    & River Blackwater & 151817001
\\
9 &Great-Corby             & River Eden       & 762505
\\
10 &Elm-Park-(Bretons-Farm) & Beam River       & 5541TH
\\
\hline
\end{tabular}
\caption{Details of the data set containing daily river flow recorded at ten different stations in the UK from 01/01/2000 to 31/12/2019.}
\label{tab:river}
\end{table}
\end{singlespace}

\begin{figure}[H]
\centering
\includegraphics[width=8cm]{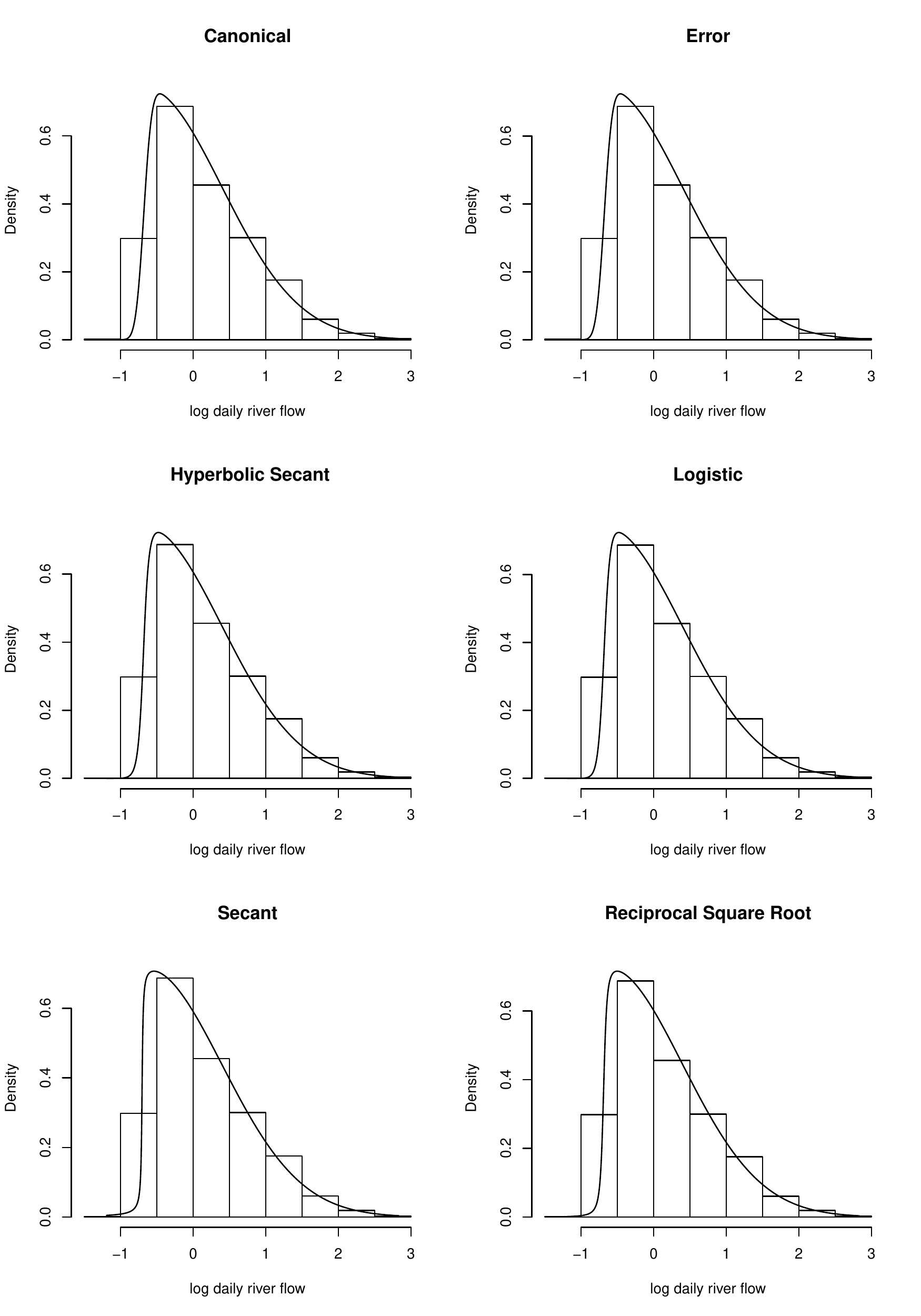}
\caption{Density plots of the fitted skew $t$-distributions on $\log$ daily river flow of river 6.}
\label{fig:river6}
\end{figure}

MLEs are obtained via the $optim$ routine in R (R Core Team, 2019).
Maximum log-likelihoods obtained and runtimes for each model and each data set are shown in Table \ref{tab:univariate}.
We can see that on average, skew $t$-distribution with error or logistic skewing function obtained
similar log-likelihood as the canonical one, but the times required to fit the model are shorter on average in both cases.
Moreover, despite the significant differences between log-likelihood obtained in the river 6 data,
the fitted pdfs as shown in Figure \ref{fig:river6} show that all the considered models can reasonably capture the shape of the pdfs.

\begin{table}[H]
\resizebox{\textwidth}{!}{
\begin{tabular}{@{}|c|cccccccccc|c|@{}}
\toprule
Skewing function / River            & 1             & 2             & 3             & 4             & 5             & 6             & 7             & 8             & 9             & 10            & Average
\\
\midrule
\multirow{2}{*}{Canonical}              & {\bf-7942.28}      & -10673.62     & -4277.54      & {\bf-10446.48}     & -7408.60      & -6123.02      & -9713.98      & {\bf-10086.99}     & {\bf-8547.90}      & -8163.15      & {\bf-8338.36}
\\
                                        & \small\textit{1.64} & \small\textit{1.73} & \small\textit{2.95} & \small\textit{1.71} & \small\textit{1.38} & \small\textit{1.20} & \small\textit{1.19} & \small\textit{0.85} & \small\textit{1.43} & \small\textit{2.85} & \small\textit{1.69}
\\
\multirow{2}{*}{Error}                  & -7942.46      & {\bf-10673.62}     & -4291.34      & -10446.49     & -7408.06      & -6123.02      & {\bf-9713.94}      & {\bf-10086.99}     & -8547.92      & {\bf-8161.22}      & -8339.51
\\
                                        & \small\textit{1.95} & \small\textit{1.12} & \small\textit{1.54} & \small\textit{1.04} & \small\textit{0.80} & \small\textit{1.49} & \small\textit{0.92} & \small\textit{0.50} & \small\textit{1.25} & \small\textit{1.98} & \small\textit{1.26}
\\
\multirow{2}{*}{Hyperbolic secant}      & -7950.39      & -10676.08     & {\bf-4271.74}      & -10451.38     & -7409.06      & -6087.13      & -9723.02      & -10098.00     & -8591.52      & -8164.42      & -8342.27
\\
                                        & \small\textit{1.03} & \small\textit{0.51} & \small\textit{1.53} & \small\textit{1.50} & \small\textit{0.89} & \small\textit{0.97} & \small\textit{0.91} & \small\textit{0.83} & \small\textit{1.21} & \small\textit{1.04} & \small\textit{1.04}
\\
\multirow{2}{*}{Logistic}               & -7947.06      & -10674.76     & -4275.32      & -10449.62     & -7408.73      & -6091.32      & -9719.47      & -10093.59     & -8574.02      & -8163.23      & -8339.71
\\
                                        & \small\textit{1.39} & \small\textit{1.08} & \small\textit{1.36} & \small\textit{0.92} & \small\textit{0.77} & \small\textit{1.12} & \small\textit{0.88} & \small\textit{1.07} & \small\textit{1.05} & \small\textit{1.75} & \small\textit{1.14}
\\
\multirow{2}{*}{Secant}                 & -8060.23      & -11050.52     & -4339.44      & -10515.02     &{\bf -7401.94}      & -6125.41      & -10065.66     & -10208.28     & -8786.11      & -8235.08      & -8478.77
\\
                                        & \small\textit{1.23} & \small\textit{0.46} & \small\textit{1.90} & \small\textit{1.56} & \small\textit{1.14} & \small\textit{1.51} & \small\textit{0.39} & \small\textit{1.56} & \small\textit{0.44} & \small\textit{0.98} & \small\textit{1.12}
\\
\multirow{2}{*}{Reciprocal square root} & -7985.23      & -10703.88     & -4280.35      & -10469.58     & -7402.95      & {\bf-6082.73}      & -9758.06      & -10131.50     & -8664.94      & -8182.76      & -8366.10
\\
                                        & \small\textit{1.35} & \small\textit{1.05} & \small\textit{1.88} & \small\textit{1.29} & \small\textit{1.24} & \small\textit{1.57} & \small\textit{1.22} & \small\textit{0.69} & \small\textit{1.07} & \small\textit{1.35} & \small\textit{1.27}
\\
\midrule
\end{tabular}}
\caption{Maximum log-likelihood of skew $t$ distributions with different skewing functions on data sets
containing logarithm of daily river flow ($m/s^2$) recorded at ten different stations in the UK from 01/01/2000 to 31/12/2019.
The runtimes (seconds) to fit each model are in {\textit {italic}}, and the highest log-likelihoods obtained for each data set are in {\bf bold}.}
\label{tab:univariate}
\end{table}

We have demonstrated that univariate canonical skew distributions are often slow in
comparison and they do not always provide better approximations then other cases of GSE.
However, the major problem of GSE is that its complexity of skewing function only grows
linearly with the number of dimensions.
For instance, the skewing function in GSE is always a univariate sigmoid function,
while canonical skew distributions such as the multivariate skew $t$ in Sahu et al. (2003)
admit more complex skewing functions in higher dimensions.
In fact, the proposed class can be viewed as a generalisation of GSE, which is capable to
admit more complex skewing functions and have GSE as a special case.

\section{Multivariate canonical skew distributions}
\label{section:multivariate}

Before we describe the proposed class, a brief review on multivariate canonical distributions is presented to
help explain the practical advantage of the proposed class.
A sensible starting point is the multivariate skew-normal distribution (Azzalini and Dalla-Valle, 1996) which has the stochastic representation
\begin{align}
\label{eq:amsn}
\displaystyle
\left[\begin{matrix}
X_0
\\
{\bf X}_1
\end{matrix}\right]
\sim N_{1+k}\left(
\left[\begin{matrix}
0
\\
{\bf 0}
\end{matrix}\right],
\left[\begin{matrix}
1 & {\bm \delta}^T
\\
{\bm \delta}  &{\bm \Omega}
\end{matrix}\right]\right),
\end{align}
where $X_0 \in \mathbb{R}$, ${\bf X}_1 \in \mathbb{R}^k$, ${\bm \delta}$ is a $k \times 1$ vector and ${\bm \Omega}$ is a $k \times k$ covariance matrix, and
\begin{align*}
\displaystyle
{\bf X} = {\bf X}_1 | X_0 > 0.
\end{align*}
It can be shown that ${\bf X}$ follows the multivariate skew-normal distribution with pdf
\begin{align*}
\displaystyle
f_{\bf X} \left({\bf x};{\bm \Omega},{\bm \delta}\right)  = 2\psi_k \left({\bf x}; {\bf 0}, {\bm \Omega}\right)
\Psi\left(\frac {{\bm \delta}^T{\bm \Omega}^{-1}}{\left(1-{\bm \delta}^T{\bm \Omega}^{-1}{\bm \delta}\right)^{1/2}}{\bf x}\right)
\end{align*}
since we have
\begin{align*}
\displaystyle
f_{\bf X}({\bf x})
&=
\displaystyle
f_{{\bf X}_1|X_0>0}({\bf x})
\\
&=
\displaystyle
\frac {f_{{\bf X}_1}({\bf x})P\left( X_0>0 | {\bf X}_1={\bf x} \right)}{P \left(X_0>0\right)}
\end{align*}
and, from conditional distribution of multivariate normal, we have
\begin{align*}
\displaystyle
X_0 | {\bf X}_1 = {\bf x} \sim N\left( {\bm \delta}^T{\bm \Omega}^{-1}{\bf x},1-{\bm \delta}^T{\bm \Omega}^{-1}{\bm \delta}\right)
\end{align*}
with
\begin{align*}
\displaystyle
f_{{\bf X}_1} \left( {\bf x}; {\bm \Omega} \right)
&=
\displaystyle
\psi_k \left({\bf x}; {\bf 0},{\bm \Omega}\right),
\\
\displaystyle
P \left(X_0>0|{\bf X}_1={\bf x}\right)
& =
\displaystyle
1 - \Psi \left( 0; {\bm \delta}^T{\bm \Omega}^{-1}{\bf x},1-{\bm \delta}^T{\bm \Omega}^{-1}{\bm \delta}\right)
\\
& =
\displaystyle
\Psi\left({\bm \delta}^T{\bm \Omega}^{-1}{\bf x}; {\bf 0}, 1-{\bm \delta}^T{\bm \Omega}^{-1}{\bm \delta}\right),
\\
\displaystyle
P\left(X_0>0\right) &= \frac {1}{2}.
\end{align*}

Many later extensions to the multivariate skew-normal distribution remain in a similar canonical form as (\ref{eq:amsn}).
As shown in  Arellano-Valle and Genton (2005), canonical skew distributions have the stochastic representation
\begin{align*}
\displaystyle
\left[\begin{matrix}
{\bf X}_0
\\
{\bf X}_1\end{matrix}\right] \sim El_{k_0+k_1}\left(
\left[\begin{matrix}
{\bm \mu}_0
\\
{\bm \mu}_1
\end{matrix}\right],
\left[\begin{matrix}
{\bm \Omega}_0 & {\bm \delta}^T
\\
{\bm \delta}  &{\bm \Omega}_1
\end{matrix}\right],
g^{\left(k_0 + k_1 \right)} \right)
\end{align*}
and
\begin{align*}
\displaystyle
{\bf X} = {\bf X}_0 | {\bf X}_1  > {\bm \tau},
\end{align*}
where $El$ is an elliptical distribution, $g^{(k)}$ is a density generator function defined
in (\ref{eq:generator}) for a $k$-dimensional elliptical distribution with kernel $g$
\begin{align}
\label{eq:generator}
\displaystyle
g^{(k)}(\mu) = \frac {\Gamma(k/2)}{\pi^{k/2}}
\frac {\displaystyle g(\mu;k)}{\displaystyle \int_0^\infty r^{k/2-1}g(r;k)dr}
\end{align}
and the pdf for ${\bf x} \in \mathbb{R}^k$ for some location parameter $\psi$ and scale parameter ${\bm \Sigma}$ is
\begin{align*}
\displaystyle
f\left({\bf x}; {\bm\psi}, {\bm \Sigma}\right) = g^{(k)}\left(\left({\bf x} - {\bm\psi} \right)^T{\bm \Sigma}^{-1}\left({\bf x} - {\bm\psi}\right)\right).
\end{align*}

Consider the pdf of ${\bf X}$, we have
\begin{align*}
\displaystyle
f_{\bf X} ({\bf x}) = \frac {f_{{\bf X}_1}({\bf x})P \left( {\bf X}_0>{\bm \tau}|{\bf X} = {\bf x} \right)}{P \left( {\bf X}_0 >{\bm \tau}\right)}.
\end{align*}
By Theorem 2.18 in Fang et al. (1990), we know that the conditional distribution of
${\bf X}_0 | {\bf X}_1 = {\bf x}_1$ remains an elliptical distribution, defined as
\begin{align}
\label{eq:conditionalg}
\displaystyle
{\bf X}_0 | {\bf X}_1 = {\bf x}_1 \sim El \left( {\bm \mu}_{0.1},{\bm \Omega}_{0.1},g_{q\left( {\bf x}_1\right)}^{\left(k_0\right)} \right),
\end{align}
where
\begin{align*}
&
\displaystyle
{\bm \mu}_{0.1} = {\bm \mu}_0 +{\bm \delta}^T{\bm \Omega}_1^{-1} \left({\bf x}_1 - {\bm \mu}_1 \right),
\\
&
\displaystyle
{\bm \Omega}_{0.1} = {\bm \Omega}_0 - {\bm \delta}^T{\bm \Omega}_1^{-1}{\bm \delta},
\\
&
\displaystyle
q\left( {\bf x}_1 \right) = \left( {\bf x}_1 - {\bm \mu}_1 \right)^T{\bm \Omega}_1^{-1} \left( {\bf x}_1 - {\bm \mu}_1 \right)
\end{align*}
and
\begin{align*}
\displaystyle
g^{\left(k_0\right)}_{q \left( {\bf x}_1 \right)}(\mu)= \frac {\Gamma \left( \left(k_0+k_1\right)/2\right)}{\pi^{k_0/2}}
\frac {\displaystyle g\left(q \left( {\bf x}_1 \right) + \mu;k_0+k_1\right)}
{\displaystyle \int_0^{\infty}r^{k_0/2-1}g\left(q \left( {\bf x}_1 \right)+r;k_0+k_1\right) dr}.
\end{align*}
Hence, the exact pdf of ${\bf X}$ can be achieved with these information.
However, a closed-form pdf for $g^{\left(k_0\right)}_{q\left( {\bf x}_1 \right)}$ does not always exist
even if $g^{(k)}$ is in closed-form.
Two of the known cases with closed-form pdfs are when $g$ is the Student's $t$ kernel or the Gaussian kernel.
When $g$ is the Student's $t$ kernel, its $g^{\left(k_0\right)}_{q\left( {\bf x}_1 \right)}$ counter part can be obtained as follows.

Consider the Student's $t$ kernel
\begin{align*}
\displaystyle
g(\mu;k,\nu) = \left(1+\frac {\mu}{\nu}\right)^{-(\nu+k)/2}.
\end{align*}
From (\ref{eq:conditionalg}), if $\left( {\bf X}_0, {\bf X}_1\right)$ follows multivariate $t$,
then the kernel pdf of $f_{{\bf X}_0|{\bf X}_1} \left({\bf x}_0|{\bf x}_1\right)$ is
\begin{align*}
\displaystyle
f_{{\bf X}_0 |{\bf X}_1}\left({\bf x}_0|{\bf x}_1\right)
&\propto
\displaystyle
\left(1+\frac {q\left({\bf x}_1\right)+\left({\bf x} - {\bm \mu}_{0.1}\right)^T{\bm \Omega}_{0.1}^{-1}
\left({\bf x} - {\bm \mu}_{0.1}\right)}{\nu}\right)^{-\left(\nu+k_0+k_1\right)/2}
\\
\displaystyle
&\propto
\left(1+\frac {\left({\bf x}-{\bm \mu}_{0.1}\right)^T{\bm \Omega}_{0.1}^{-1}
\left({\bf x}-{\bm \mu}_{0.1}\right)}{\nu+q\left({\bf x}_1\right)}\right)^{-\left(\nu+k_0+k_1\right)/2}
\\
\displaystyle
&\propto
\left(1+\frac {\left({\bf x}-{\bm \mu}_{0.1}\right)^T
{\bm \Omega}_{0.1}^{-1}\left({\bf x}-{\bm \mu}_{0.1}\right)}{\nu+k_1}\frac {\nu+k_1}{\nu+q \left({\bf x}_1\right)}\right)^{-\left(\nu+k_0+k_1\right)/2}
\\
\displaystyle
&\propto
\left(1+\frac {\left({\bf x}-{\bm \mu}_{0.1}\right)^T
\left(\frac {\nu+q\left({\bf x}_1\right)}{\nu+k_1}{\bm \Omega}_{0.1}\right)^{-1}
\left({\bf x}-{\bm \mu}_{0.1}\right)}{\nu+k_1}\right)^{-\left(\nu+k_0+k_1\right)/2}.
\end{align*}
Hence, ${\bf X}_0|{\bf X}_1 \sim t_{k_0}\left({\bm \mu}_{0.1}, \frac {\nu+q\left({\bf x}_1\right)}{\nu+k_1}{\bm \Omega}_{0.1}, \nu+k_1\right)$,
and therefore, the pdf of ${\bf X}$ is
\begin{align}
\label{eq:mst}
\displaystyle
f_{{\bf X}}\left({\bf x};{\bm \mu}, {\bm \Omega}, {\bm \tau}\right) =
\frac {\displaystyle t_{k_1}\left( {\bf x}-{\bm \mu}_1;  {\bf 0}, {\bm \Omega}_1,\nu\right)
T_{k_0}\left(\sqrt{\frac {\nu+k_1}{\nu+q \left({\bf x}\right)}}\left({\bm \mu}_{0.1} - {\bm\tau} \right);  {\bf 0}, {\bm \Omega}_{0.1},\nu+k_1\right)}
{\displaystyle T_{k_0}\left( {\bm \mu}_{0} - {\bm\tau};  {\bf 0}, {\bm \Omega}_{0},\nu\right)},
\end{align}
where $t_{k}$ and $T_{k}$ denote, respectively, the joint pdf and joint cdf of a $k$-dimensional multivariate skew-$t$ distribution.

Many existing multivariate skew distributions can be represented as special cases of (\ref{eq:mst}) under different parameterisations.
Multivariate skew normal distributions in Azzalini and Dalla Valle (1996), Branco and Dey (2001), Lachos et al. (2010),
and Pyne et al. (2009) took the nested form of (\ref{eq:mst}) with $k_0 = 1$, ${\bm \tau} = {\bm \mu}_0$, and $\nu = \infty$,
and the denominator in (\ref{eq:mst}) becomes $\frac {1}{2}$;
Sahu et al. (2003) took $k_0 = k_1$, ${\bm \tau} = {\bm \mu}_0$, $\nu = \infty$, and ${\bm \Omega}_0 = {\bf I}_{k_1}$,
and the denominator becomes $2^{-k_1}$;
Gupta et al. (2014) took $k_0 = k_1$, $\nu = \infty$, and ${\bm \tau} = {\bm \mu}_0$;
Azzalini and Capitanio (1999) took $k_0 = 1$ and $\nu = \infty$;
Arellano-Valle and Genton (2005) took ${\bm \tau} = {\bm \mu}_0$, $\nu = \infty$, ${\bm \Omega}_0 = {\bf I}_{k_0}$,
and  ${\bm \Omega}_1 = {\bf I}_{k_1}$, and their distribution named the canonical fundamental skew normal distribution (CFUSN);
Gonzalez-Faras et al. (2004), Liseo and Loperfido (2003), and Arellano-Valle and Azzalini (2006) simply took $\nu = \infty$,
resulting in the most general form of canonical skew normal distributions.

Similarly, when $\nu$ is finite, the distribution of ${\bf X}$ remains as multivariate skew $t$.
Branco and Dey (2001), Azzalini and Capitanio (2003), Gupta (2003), Pyne et al. (2009), and
Lacho et al. (2010) took $k_0 = 1$, and ${\bm \tau} = {\bm \mu}_0$, the denominator becomes $\frac {1}{2}$;
Sahu et al. (2003) took $k_0 = k_1$, ${\bm \Omega}_0 = {\bf I}_{k_0}$, and ${\bm \tau} = {\bm \mu}_0$,
the denominator becomes $2^{-k_1}$;
Arellano-Valle and Genton (2005) took ${\bm \tau} = {\bm \mu}_0$, ${\bm \Omega}_0 = {\bf I}_{k_0}$,
and ${\bm \Omega}_1 = {\bf I}_{k_1}$, and their distribution named the canonical fundamental skew $t$ distribution (CFUST);
Arellano-Valle and Genton (2010a) developed a multivariate skew $t$ distribution similar to (\ref{eq:mst}) with different parameterisation.

As explained, the stochastic structure allows researchers to understand better the
statistical properties of canonical skew distributions.
For example, Arellano-Valle and Genton (2005) proved that the moment generating function of the CFUSN can be expressed analytically as
\begin{align*}
\displaystyle
M_{\bf X} ({\bf t}) = 2^{k_1}\exp\left(1/2\right) {\bf t}^T {\bf t} \Phi_{k_1}\left({\bm \delta}^T {\bf t}\right)
\end{align*}
and the joint cdf can also be expressed analytically as
\begin{align*}
\displaystyle
M_{\bf X}({\bf x}) = 2^{k_1} \Phi_{k_0+k_1}\left(\left({\bf x}^T, {\bf 0}^T\right)^T | {\bm \Sigma} \right),
\end{align*}
where ${\bm \Sigma} = \left[\begin{matrix}
{\bf I}_{k_0} & -{\bm \delta}^T
\\
-{\bm \delta}  &{\bf I}_{k_1}
\end{matrix}
\right]$.
Moreover,  the CFUSN  allows closed-form Markov Chain Monte Carlo (MCMC) for Bayesian inference.
For example, Sahu et al. (2003) developed a MCMC algorithm for Bayesian inference of regression with
multivariate skew normal distributions;
Fruhwirth-Schnatter and Pyne (2010) developed a MCMC algorithm for finite mixtures of both skew-normal and skew-$t$ distributions;
Liseo and Parisi (2013) developed a Bayesian method for skew-normal distributions.

Although canonical skew-$t$ and skew-normal distributions have been studied extensively,
inference for these distributions is not an easy task in practice.
When $k_1$ is moderate, say, $k_1 > 5$, for instance.
Since closed-form estimators do not exist in general for this class,
MLEs require the evaluation of $T_{k}$ or $\Phi_{k}$, which can be computationally intensive.
In some special cases, evaluation of $\Phi_{k}$ can be omitted under some hierarchical MCMC setups for Bayesian inference,
but there is yet to be an efficient MCMC algorithm for the general form (\ref{eq:mst}).
Flexible multivariate distributions, moreover, are often used in mixture modelling,
in which the computational cost would be further magnified.
In view of this, we aim to develop a new class of multivariate skew distributions as flexible as (\ref{eq:mst}),
while being computationally less heavy to fit.

Arellano-Valle and Genton (2005) generalised a class of skew distributions
called the fundamental skew distributions, they described a general method
to introduce skewness into distributions via perturbation.
For instance, consider a $k$-variate elliptical distribution $f_k$, a skew-elliptical distribution
can be obtained with some skewing function $g_m$ such that the resulting pdf is
\begin{align}
\label{eq:transform_s}
\displaystyle
f_k\left({\bf x}|g_m\right) = \frac {\displaystyle g_m \left({\bf x}\right) f_k \left({\bf x}\right)}{\displaystyle \int_{\mathbb{R}^k}g_m \left({\bf u}\right)f_k \left({\bf u}\right)d{\bf u}}.
\end{align}

Clearly, (\ref{eq:mst}) belongs to this class of distributions,
by having $f_k$ and $g_m$ as the pdf and cdf of multivariate $t$ of appropriate degrees of freedom.
The stochastic representation of (\ref{eq:transform_s}) can also be described as a hidden truncation where
\begin{align*}
\displaystyle
{\bf X} = {\bf X}_0 | {\bf X}_1 > {\bf 0},
\end{align*}
where $f_{{\bf X}_0}\left( {\bf x}_0\right) = f_k\left({\bf x}_0\right)$ and
$P\left({\bf X}_1 > {\bf 0} | {\bf X}_0 = {\bf x}_0\right) = g_m\left(  {\bf x}_0\right)$.

When $g_m$ is not the canonical counter part of $f_k$, the integral in (\ref{eq:transform_s}) is generally not analytical,
which can create another problem for statistical inference.
Finally, we introduce the proposed class of distributions in next section,
and we show that sacrificing the analyticity of pdf can be beneficial in some cases.

\section{A novel class of skew distributions}
\label{section:gms}

We call the proposed class of distributions the skew elliptical distributions with independent skewing functions (SELIS).
Its pdf has the following form
\begin{align}
\label{eq:gen_s}
\displaystyle
f_k\left({\bf x};  {\bm\mu}, {\bf A}, \nu, {\bm \lambda}, g_m \right)
\displaystyle
&=\frac {\displaystyle g_m\left( {\bm \lambda} {\bf A}^{-1}\left({\bf x} - {\bm\mu} \right)\right)
El\left({\bf x};  {\bm\mu}, {\bf A} {\bf A}^{T},\nu\right)}
{\displaystyle \int_{{\bf u} \in \mathbb{R}^k}{g_m\left( {\bm \lambda} {\bf A}^{-1}
\left( {\bf u} - {\bm\mu}  \right)\right)
El\left( {\bf u};  {\bm\mu}, {\bf A} {\bf A}^{T}, \nu\right) d\bf u}}
\nonumber
\\
\displaystyle
&= \frac {\displaystyle g_m\left( {\bm \lambda} {\bf A}^{-1}\left( {\bf x} - {\bm\mu} \right)\right)
El\left({\bf x};  {\bm\mu}, {\bf A} {\bf A}^{T}, \nu\right)}
{\displaystyle \int_{{\bf v} \in \mathbb{R}^k}{g_m\left( {\bm \lambda} {\bf v} \right)}{El\left({\bf v}; {\bf 0}, {\bf I}_k, \nu\right)}d{\bf v}}
\end{align}
and
\begin{align*}
\displaystyle
g_m \left( {\bm \lambda} {\bf v} \right) = \prod_{i=1}^m g\left( {\bm \lambda}_{i} {\bf v} \right),
\end{align*}
where $\bm\mu$ is a $k$ dimensional vector, ${\bf A}$ is a $k \times k$ lower triangular matrix such
that ${\bf A} {\bf A}^T$ represents the scale matrix, $\nu$ is the shape parameter of the elliptical distribution,
${\bm \lambda}$ is a $m \times k$ upper triangular skewing matrix, and $g$ is some sigmoid function.

Note that ${\bm \lambda}$ is an upper triangular matrix instead of a full matrix because
any full matrix can be decomposed by any orthogonal matrix ${\bf Q}$ as ${\bf RQ}$,
where ${\bf R}$ is an upper triangular matrix.
We write ${\bm \lambda} = {\bf RQ}$, and ${\bf B}^{-1} = {\bf Q} {\bf A}^{-1}$.
Clearly, ${\bf A} {\bf A}^{T} = {\bf B} {\bf B}^{T}$ and therefore
\begin{align*}
&\phantom{= }
\displaystyle
g_m\left( {\bm \lambda} {\bf A}^{-1}\left( {\bf x} - {\bm\mu}\right)\right) El\left( {\bf x};  {\bm\mu}, {\bf A} {\bf A}^{T},\nu\right)
\\
\displaystyle
&= g_m\left( {\bf RQ} {\bf A}^{-1}\left( {\bf x} - {\bm\mu} \right)\right) El\left( {\bf x};  {\bm\mu}, {\bf A} {\bf A}^{T},\nu\right)
\\
\displaystyle
&= g_m\left( {\bf R} {\bf B}^{-1}\left( {\bf x} - {\bm\mu} \right)\right) El\left( {\bf x};  {\bm\mu}, {\bf B} {\bf B}^{T},\nu\right).
\end{align*}
The pdf is not unique when ${\bm \lambda}$ is a full matrix.

The integral in (\ref{eq:gen_s}) does not have an analytical expression.
However, it can be written as the expected value of $g_m\left( {\bm \lambda} {\bf Y}\right)$ with ${\bf Y}$ following some $k$-variate
elliptical distribution of location vector ${\bf 0}$, scale matrix ${\bf I}_k$, and shape parameter $\nu$.
Under some classes of elliptical distributions, random variables can be generated efficiently.
Multivariate normal, multivariate $t$, and multivariate power exponential (Gomez, 1998) are examples.
Hence, the integral can be estimated reasonably efficiently by Monte carlo integration.

The skewing function $g_m$ as opposed to the canonical skew $t$ distribution in (\ref{eq:mst})
is a product of $m$ sigmoid where the computational cost is a linear function of $m$.
When $m$ is large it would be significantly more efficient to evaluate $g_m$ than
to evaluate the $m$-variate joint cdf in (\ref{eq:mst}) of same complexity.
Although the pdf in (\ref{eq:gen_s}) is non-analytical, it can be expressed in closed-form up to proportionality.
In other words, the non-analytical part only needs to be evaluated once for each set of parameters.
(\ref{eq:mst}), on the other hand, requires the evaluation of joint cdf for each data point.
Therefore, evaluations of log-likelihood of (\ref{eq:gen_s}) should be significantly faster, especially when the data set is large.

Some canonical skew distributions  can be obtained via convolution.
The SELIS generally can not be expressed in terms of convolution of known distributions,
one possible method to simulate random variables of  the SELIS is by the conditional method:
\begin{enumerate}

\item
draw sample ${\bf x}$ from the appropriate elliptical distribution;

\item
draw samples $u$ from $U(0,1)$;

\item
accept $x$ if $u < g_m\left( {\bm \lambda} {\bf A}^{-1}\left( {\bf x} - {\bm \mu} \right)\right)$;

\item
repeat steps 1 to 3 until a desirable number of samples is obtained.

\end{enumerate}
However, it is not an efficient method to generate random variables.
Since the expected acceptance rate is $E_{\bf U} \left[g_m \left( {\bm \lambda} {\bf u} \right) \right]$,
where ${\bf U}$ is spherically distributed with an appropriate generator,
$E_{\bf U} \left[g_m \left( {\bm \lambda} {\bf u} \right) \right]$ decreases exponentially with $m$.

\subsection{Special cases}

Several existing multivariate skew distributions can be considered special cases of  the SELIS.
For instance, when $m = 1$,  the SELIS reduces to GSE, which contains many existing multivariate skew distributions.
Also, another special case can be achieved when multivariate normal is considered in (\ref{eq:mst}),
i.e. setting $\nu = \infty$.
The canonical skew normal distribution can be written in form of (\ref{eq:gen_s})
if ${\bm \Omega}_{0.1} = {\bm \Omega}_0 - {\bm \delta}^T{\bm \Omega}_1^{-1}{\bm \delta} = {\bf I}$.
In that case, the skewing function is the joint cdf of independent normal distributions,
which can be written in the form of $g_m$.
There does not appear to be other cases of canonical skew distributions
that can be written in form of (\ref{eq:gen_s}) since non-Gaussian elliptical distributions are generally not independent, even when they are uncorrelated.

Moreover, there exist cases when the integral in (\ref{eq:gen_s}) is known.
For instance, when the skewing matrix is a square and diagonal.
In that case, the integral in (\ref{eq:gen_s}) becomes $2^{-k}$.
A simple proof is as follows.

When ${\bm \lambda}$ is diagonal, the integral in (\ref{eq:gen_s}) can be written as
\begin{align*}
\displaystyle
\underset{k}{\int_{-\infty}^{\infty}\cdots\int_{-\infty}^{\infty}}
\left[\prod_{i=1}^k\left(g_k\left(\lambda_{ii}x_i\right)\right) \right]
El \left( {\bf x}^{(k)}, {\bf 0}, {\bf I}_k, \nu \right) dx_1 \cdots dx_k.
\end{align*}
Then, we can show
\begin{align}
\phantom{=}
\displaystyle
&\underset{k}{\int_{-\infty}^{\infty}\cdots\int_{-\infty}^{\infty}}
\left[\prod_{i=1}^kg_k \left(\lambda_{ii}\right) \right]
El\left( {\bf x}^{(k)}, {\bf 0}, {\bf I}_k, \nu \right) dx_1 \cdots dx_k
\nonumber
\\
\displaystyle
&= \underset{k-1}{\int_{-\infty}^{\infty}\cdots\int_{-\infty}^{\infty}}
\int_{0}^{\infty}g_k \left(\lambda_{11}\right)
\left[\prod_{i=2}^kg_k \left(\lambda_{ii}\right) \right]
El \left( {\bf x}^{(k)}, {\bf 0}, {\bf I}_{k}, \nu \right) dx_1 \cdots dx_k
\nonumber
\\
\displaystyle
&
\quad
+ \underset{k-1}{\int_{-\infty}^{\infty}\cdots\int_{-\infty}^{\infty}}
\int_{0}^{\infty}\left(1 - g_k \left(\lambda_{11}\right)\right)
\left[\prod_{i=2}^k g_k \left(\lambda_{ii}\right) \right]
El \left( {\bf x}^{(k)}, {\bf 0}, {\bf I}_{k}, \nu \right) dx_1 \cdots dx_k
\nonumber
\\
\displaystyle
& = \underset{k-1}{\int_{-\infty}^{\infty}\cdots\int_{-\infty}^{\infty}}
\int_{0}^{\infty}\left[\prod_{i=2}^k g_k \left(\lambda_{ii}\right) \right]
El \left( {\bf x}^{(k)}, {\bf 0}, {\bf I}_{k}, \nu \right) dx_1 \cdots dx_k
\nonumber
\\
\displaystyle
& = \frac {1}{2} \underset{k-1}{\int_{-\infty}^{\infty}\cdots
\int_{-\infty}^{\infty}}
\left[\prod_{i=2}^k g_k \left(\lambda_{ii}\right) \right]
El \left( {\bf x}^{(k-1)}, {\bf 0}, {\bf I}_{k-1}, \nu \right) dx_2 \cdots dx_k
\nonumber
\\
\displaystyle
& = 2^{-k}.
\label{eq:diag_proof}
\end{align}

Extension of (\ref{eq:diag_proof}) for $m \times k$ diagonal skewing matrices
can be easily obtained by adding or removing rows of zeros in from the diagonal square ${\bm \lambda}$.
In that case, (\ref{eq:gen_s}) becomes $2^{-m}$.
Moreover, it can be shown that for any $m \times k$ skewing matrix ${\bm \lambda}$ the maximum of the integral is $2^{-m}$.
By H\"{o}lder's inequality,
\begin{align*}
\displaystyle
E_{\bf U}\left[g_m\left( {\bm \lambda} {\bf u}\right) \right]
&=
\displaystyle
E_{\bf U} \left[  \left | g_m\left( {\bm \lambda} {\bf u} \right) \right | \right]
\\
& \le
\displaystyle
\prod_{i=1}^m E_{\bf U} \left[ \left | g\left( {\bm \lambda}_i {\bf u} \right) \right | \right]
\\
& =
\displaystyle
\prod_{i=1}^m E_{\bf U} \left[ g\left( {\bm \lambda}_i {\bf u} \right) \right]
\\
& =
\displaystyle
2^{-m}.
\end{align*}

\subsection{Statistical inference}
\label{section:inference}

Without analytical pdfs, obtaining ML estimates is difficult and computationally intensive
if brute force numerical optimisation is to be applied.
Also, since  the SELIS does not have simple convolution type stochastic representation as some canonical skew distributions do,
closed-form MCMC method is also unavailable.
In this section, we investigate some possible inference methods to obtain ML estimates of  the SELIS via gradient based methods.
We show that  the SELIS can be fitted within reasonable time frame with the proposed algorithm.

Consider the log-likelihood function of  the SELIS
\begin{align}
\displaystyle
\pi\left( {\bm \mu}, {\bf A}, \nu, \lambda | {\bf x} \right)
&=
\displaystyle
\sum_{i=1}^n\left[\log\left(g\left(\lambda {\bf A}^{-1} \left( {\bf x}_i - {\bm\mu} \right)\right)\right)
+ \log\left(El \left( {\bf x}_i;  {\bm \mu}, {\bf A} {\bf A}^T, \nu \right) \right) \right]
\nonumber
\\
&
\displaystyle
\quad
-n \log \int_{{\bf u} \in \mathbb{R}^k}
g\left(\lambda {\bf u} \right)
El \left( {\bf u};  {\bf 0}, {\bf I}_k,\nu\right) d {\bf u},
\label{eq:logl}
\end{align}
where ${\bm\mu}$ denotes the location vector, ${bf A}$ denotes some lower triangular matrix
such that ${\bf A} {\bf A}^T$ denotes the scale matrix of elliptical distributions, and $\nu$ denotes the shape parameter.

Consider the gradients of the integral part in (\ref{eq:logl}) with respect to $\lambda$ and $\nu$, we have
\begin{align}
\displaystyle
\frac {\displaystyle d\log \int_{{\bf u} \in \mathbb{R}^k}
g_m\left(\lambda {\bf u} \right)El\left({\bf u};  {\bf 0}, {\bf I}_k,\nu\right) d{\bf u}}{\displaystyle d\lambda}
&=
\displaystyle
\frac {\displaystyle \int_{{\bf u} \in \mathbb{R}^k}
\frac {d\log g_m\left(\lambda {\bf u}\right)}{d\lambda}g_m\left(\lambda {\bf u}\right)
El \left( {\bf u};  {\bf 0}, {\bf I}_k, \nu \right) d{\bf u}}
{\displaystyle \int_{{\bf u} \in \mathbb{R}^k}g_m\left(\lambda {\bf u} \right)El \left({\bf u};  {\bf 0}, {\bf I}_k,\nu\right) d{\bf u}}
\nonumber
\\
&=
\frac {\displaystyle E_{\bf U} \left( \frac {d\log g_m\left(\lambda {\bf u} \right)}
{d\lambda}g_m\left(\lambda {\bf u} \right) \right)}{\displaystyle E_{\bf U} \left( g_m\left(\lambda {\bf u} \right) \right)},
\label{eq:dlambda}
\end{align}
and
\begin{align}
\displaystyle
\frac {\displaystyle d\log \int_{{\bf u} \in \mathbb{R}^k}g_m\left(\lambda {\bf u}\right)
El \left({\bf u};  {\bf 0}, {\bf I}_k,\nu\right) d{\bf u}}{\displaystyle d\nu}
&=
\displaystyle
\frac {\displaystyle \int_{{\bf u} \in \mathbb{R}^k} \frac {d\log El \left({\bf u};  {\bf 0}, {\bf I}_k, \nu\right)}{d\nu}
g_m\left(\lambda {\bf u} \right)El \left({\bf u};  {\bf 0}, {\bf I}_k,\nu\right) d{\bf u}}
{\displaystyle \int_{{\bf u} \in \mathbb{R}^k}g_m\left(\lambda {\bf u} \right)El \left( {\bf u};  {\bf 0}, {\bf I}_k,\nu\right) d{\bf u}}
\nonumber
\\
&=
\displaystyle
\frac {\displaystyle E_{\bf U} \left(\frac {d\log El \left( {\bf u};  {\bf 0}, {\bf I}_k, \nu \right)}{d\nu} g_m\left(\lambda {\bf u}\right)\right)}
{\displaystyle E_{\bf U} \left(g_m\left(\lambda {\bf u} \right) \right)},
\label{eq:dnu}
\end{align}
where ${\bf U}$ is a random variable with pdf $El \left( {\bf u};  {\bf 0}, {\bf I}_k, \nu \right)$.

Other gradients of (\ref{eq:logl}) are usually expressible in closed-form if $El$
is some common elliptical distribution such as the multivariate $t$ or the multivariate power exponential distribution,
and $g$ is some common sigmoid function such as the error function or the logistic function.

For example, consider the pdf of the multivariate-$t$ distribution
\begin{align}
\displaystyle
f_k\left( {\bf x}; {\bm\mu}, {\bf A} {\bf A}^T, \nu\right) =
\frac {\Gamma \left( (\nu+k)/2\right)}{\Gamma(\nu/2)\nu^{k/2}\pi^{k/2} \left | {\bf A} \right |}
\left[1+\frac {1}{\nu}\left( {\bf A}^{-1} \left({\bf x}-{\bm \mu}\right)\right)^T
\left( {\bf A}^{-1} \left({\bf x}-{\bm \mu}\right)\right)\right]^{-(\nu+k)/2}.
\label{eq:t}
\end{align}
The gradients of (\ref{eq:t}) with respect to $\bm\mu$, ${\bf A}^{-1}$ and $\nu$ are
\begin{equation*}
\begin{aligned}
\displaystyle
\frac {d\log f_k \left( {\bf x};{\bm \mu},{\bm \Sigma}, \nu \right)}{d{\bm \mu}} =
\frac {-(\nu+k)}{\nu}\left[1+\frac {1}{\nu}\left( {\bf A}^{-1} \left({\bf x}-{\bm \mu}\right)\right)^T
\left( {\bf A}^{-1} \left({\bf x}-{\bm \mu}\right)\right)\right]^{-1} \left( {\bf A} {\bf A}^T\right)^{-1} \left({\bf x}-{\bm \mu}\right),
\end{aligned}
\end{equation*}
\begin{equation*}
\begin{aligned}
\displaystyle
\frac {d\log f_k \left( {\bf x};{\bm \mu},{\bm \Sigma},\nu\right)}{d{\bf A}^{-1}}
&=
\displaystyle
\left( {\bf A}^{-1} \circ {\bf I}\right)^{-1} +
\frac {\nu+k}{\nu}\left[1+\frac {1}{\nu}\left( {\bf A}^{-1}\left({\bf x}-{\bm \mu}\right)\right)^T
\left( {\bf A}^{-1}\left({\bf x}-{\bm \mu}\right)\right)\right]^{-1}
\\
&
\qquad
\cdot
\displaystyle
\left[ {\bf A}^{-1} \left({\bf x}-{\bm \mu}\right) \left({\bf x}-{\bm \mu}\right)^T\right]
\end{aligned}
\end{equation*}
and
\begin{equation*}
\begin{aligned}
&
\displaystyle
\frac {d\log f_k \left( {\bf x};{\bm \mu},{\bm \Sigma},\nu\right)}{d\nu} =
\displaystyle
\frac {\xi\left((\nu+k)/2\right) - \xi \left(\nu/2\right)}{2}-\frac {k}{2\nu}
\\
&
\displaystyle
\qquad
+\frac {\nu+k}{2\nu^2}\left[1+\frac {1}{\nu}\left( {\bf A}^{-1} \left({\bf x}-{\bm \mu}\right)\right)^T
\left( {\bf A}^{-1} \left({\bf x}-{\bm \mu}\right)\right)\right]^{-1}
\left( {\bf A}^{-1} \left({\bf x}-{\bm \mu}\right)\right)^T
\left( {\bf A}^{-1} \left({\bf x}-{\bm \mu}\right)\right)
\\
&
\displaystyle
\qquad
-\frac {1}{2}\log\left[1+\frac {1}{\nu}\left( {\bf A}^{-1} \left({\bf x}-{\bm \mu}\right)\right)^T
\left( {\bf A}^{-1} \left({\bf x}-{\bm \mu}\right) \right)\right],
\end{aligned}
\end{equation*}
where $\xi$ denotes the digamma function.

Consider the pdf of the multivariate power exponential distribution
\begin{align*}
\displaystyle
f_k \left( {\bf x}; {\bm\mu}, {\bf A} {\bf A}^{T}, \beta\right) =
\frac {k\Gamma(k/2)}{\pi^{k/2}{|{\bf A}|}
\Gamma\left(1+k/2\beta\right)2^{1+k/2\beta}}
\exp\left[-\frac {1}{2}\left(\left( {\bf A}^{-1}\left({\bf x}-{\bm \mu}\right)\right)^T
\left( {\bf A}^{-1} \left({\bf x}-{\bm \mu}\right)\right)\right)^\beta\right].
\end{align*}
The gradients of (\ref{eq:t}) with respect to $\bm\mu$, ${\bf A}^{-1}$ and $\beta$ are
\begin{equation*}
\begin{aligned}
\displaystyle
\frac {d\log f_k \left( {\bf x}; {\bm \mu}, {\bf A} {\bf A}^{T}, \beta\right)}{d{\bm \mu}} =
\displaystyle
\beta\left[\left( {\bf A}^{-1} \left({\bf x}-{\bm \mu}\right)\right)^T
\left( {\bf A}^{-1} \left({\bf x}-{\bm \mu}\right) \right)\right]^{\beta-1} \left( {\bf A} {\bf A}^{T}\right)^{-1} \left({\bf x}-{\bm \mu}\right),
\end{aligned}
\end{equation*}
\begin{equation*}
\begin{aligned}
\displaystyle
\frac {d\log f_k \left( {\bf x}; {\bm \mu}, {\bf A} {\bf A}^{T}, \beta \right)}{d {\bf A}^{-1}}
&=
\displaystyle
\left( {\bf A}^{-1} \circ {\bf I}\right)^{-1}+{\beta}
\left[\left( {\bf A}^{-1} \left({\bf x}-{\bm \mu}\right)\right)^T\left( {\bf A}^{-1} \left({\bf x}-{\bm \mu}\right)\right)\right]^{\beta-1}
\\
\displaystyle
&
\qquad
\cdot \left[ {\bf A}^{-1} \left({\bf x}-{\bm \mu}\right) \left({\bf x}-{\bm \mu}\right)^T\right]
\end{aligned}
\end{equation*}
and
\begin{equation*}
\begin{aligned}
&
\displaystyle
\frac {d\log f_k \left( {\bf x};{\bm \mu}, {\bm \Sigma}, \beta \right)}{d\beta} =
\frac {k}{2\beta^2} \left[ \xi \left(1+k/2\beta\right)+ \log(2) \right]
\\
&
\displaystyle
\qquad
-\frac {1}{2}\log\left[\left( {\bf A}^{-1} \left({\bf x}-{\bm \mu}\right)\right)^T
\left( {\bf A}^{-1} \left({\bf x}-{\bm \mu}\right)\right)\right]
\left[\left( {\bf A}^{-1} \left({\bf x}-{\bm \mu}\right)\right)^T\left( {\bf A}^{-1} \left({\bf x}-{\bm \mu}\right)\right)\right]^{\beta}.
\end{aligned}
\end{equation*}

Note that we use the derivative with respect to ${\bf A}^{-1}$ instead of  ${\bm \Sigma}$ or ${\bf A}$.
The advantage of this parameterisation is that both the pdf and all the gradients
can be evaluated without any matrix inversion, which can be inefficient and numerically unstable when $k$ is large.
Also, when ${\bf A}$ is a lower triangular matrix, ${\bf A} {\bf A}^T$ is guaranteed to be symmetric positive semi-definite.
This can avoid problems of scale matrix being non-positive definite during numerical optimisation.

With (\ref{eq:dlambda}) and (\ref{eq:dnu}) expressible in terms of expectation of spherical distributions,
they can be estimated systematically.
Next, we illustrate how these estimated gradients can be incorporated for MLE.

\subsubsection{MLE via gradient descent}
\label{section:sgd}

Since some gradients are not analytical, many standard deterministic numerical optimisation methods can not be used directly.
However, it is possible to incorporate the estimated gradients in first order methods such as gradient descent.
This can be viewed as a special case of stochastic gradient descent (SGD),
where the efficiency is controlled by both Monte carlo sample size and batch sample size.
An example routine for SGD is as follows:
\begin{enumerate}

\item
set $t = 0$, initial parameter ${\bm \Omega}^{(0)}$, step size $s$, batch size $n$, and Monte carlo sample size $m$;

\item
draw $n$ samples from the data set and generate $m$ Monte carlo samples to estimate the gradients of the log-likelihood function $g$;

\item
update ${\bm \Omega}^{(t+1)} = {\bm \Omega}^{(t)} + s g$;

\item
set $t = t + 1$;

\item
repeat steps  1 to 4 for a fixed number of iterations or until the convergence conditions are satisfied.

\end{enumerate}

To illustrate the efficiency of SGD for MLE of  the SELIS,
we tested the algorithm on the first two variables of the river flow data set described in Table \ref{tab:river}.
SELIS with multivariate $t$, logistic skewing function and square skewing matrix (GMST-Logistic, hereafter), and $m = 2$ was used,
and SGD with full batch size, $s = 0.01$ and $m = 10000$ was used.
For comparison, we also fitted the dataset with the multivariate skew $t$ distribution (AMST, hereafter) in Azzalini (2003).

\begin{figure}[H]
\includegraphics[width=15cm]{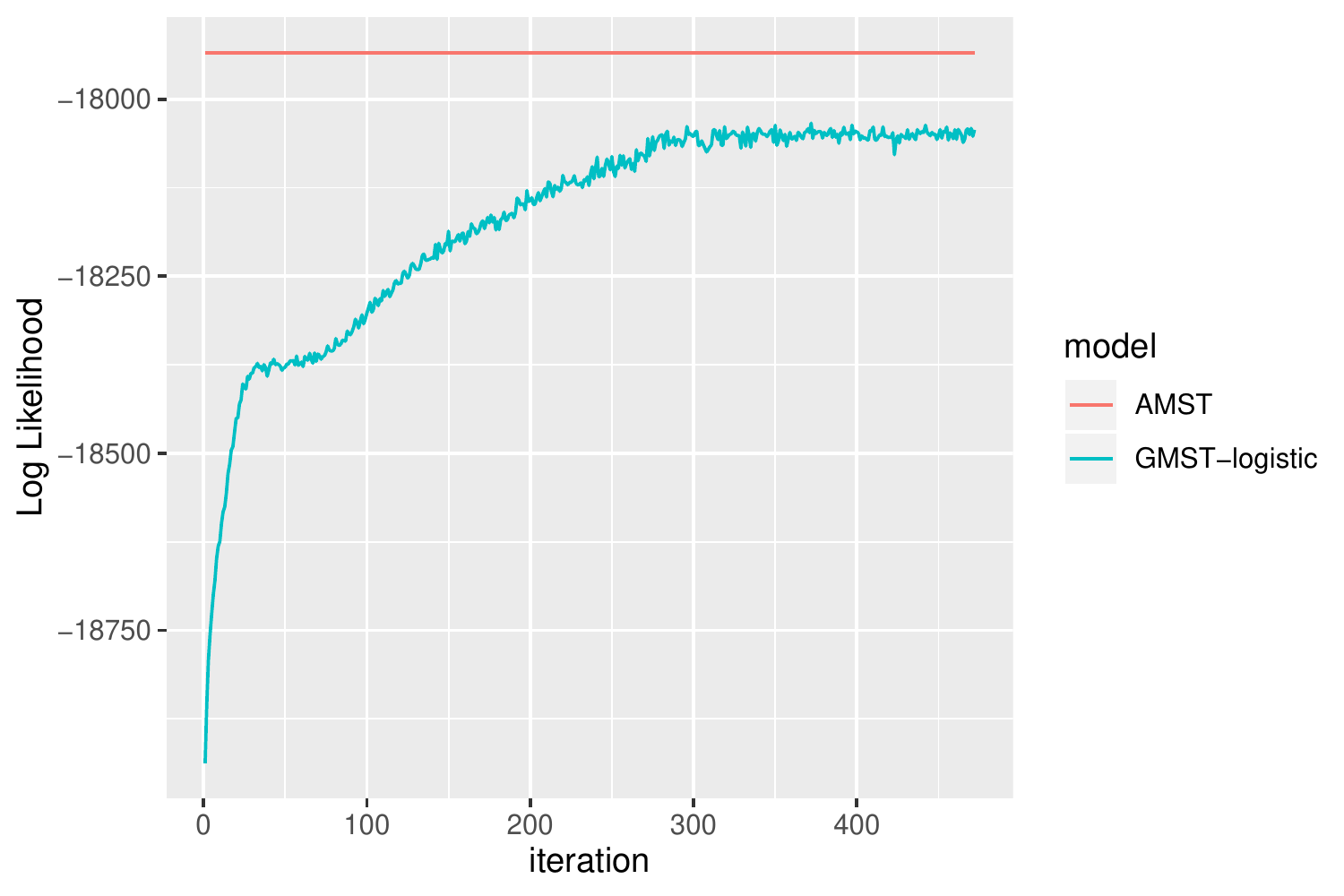}
\centering
\caption{Log-likelihood plot against iteration via SGD with 10000 Monte carlo samples,
on the first two variables of the river flow data set described in Table \ref{tab:river}.}
\label{fig:sgd}
\end{figure}

Log-likelihood at each iteration are plotted in Figure \ref{fig:sgd}.
Although the algorithm appeared to have converged after around 300 iterations,
the log-likelihood achieved is significantly lower than that obtained by the less complex AMST.
This suggests that the algorithm was stuck at a local optimum.
Possible improvements can be made if the SGD is well tuned, for example using smaller batch sizes would
allow the log-likelihood surface to be explored more efficiently.
However, when a smaller batch size is used, this SGD can not take full advantage of the gain
in computation speed as normal SGD does due to the Monte carlo integration.
And if the Monte carlo sample size is set too small, the gradients generally become too unstable for SGD to converge.
We do not aim to discuss further on tuning SGD in this paper, instead we propose a quasi-maximum likelihood method (QMLE)
that takes advantages of second order optimisation and is shown to be more efficient than SGD.

\subsubsection{Quasi-maximum likelihood estimation}

A major problem of first order methods is that they do not incorporate curvature information, hence, hindering their efficiency.
There exists some efficient stochastic quasi-Newton methods in literature such as Byrd et al. (2016)
and Moritz et al. (2016), we believe they may be suitable to find ML estimates for  the SELIS.
However, efficient implementation of these stochastic quasi-Newton methods is not a simple task,
and is generally beyond the scope of this paper.
Alternatively, we propose a QMLE method that allows the
use of deterministic quasi-Newton methods which are readily available and efficiently
implemented across many different programming languages.

Note that the distribution of the Monte carlo samples used to estimate the log-likelihood function is
dependent only on the shape parameter.
For a fixed shape parameter, the same set of Monte carlo samples can be used to estimate the log-likelihood
function with different skewing functions, location vectors, and scale matrices.
In other words, we can approximate the log-likelihood function given some fixed shape parameter
with a deterministic quasi-log-likelihood function using a fixed set of  Monte carlo samples.
The proposed quasi-log-likelihood function is as follows:
\begin{align}
\displaystyle
\pi\left( {\bm \mu}, {\bf A}, \nu, \lambda | {\bf x} \right)
&=
\displaystyle
\sum_{i=1}^n\left[\log\left(g_m\left(\lambda {\bf A}^{-1} \left( {\bf x}_i - {\bm \mu} \right)\right)\right)
+ \log\left(El \left( {\bf x}_i;  {\bm \mu}, {\bf A} {\bf A}^T, \nu \right)\right) \right]
\nonumber
\\
&
\displaystyle
\qquad
-n \log \int_{{\bf u} \in \mathbb{R}^k}
g_m\left(\lambda {\bf u}\right)El \left( {\bf u};  {\bf 0}, {\bf I}_k,\nu\right) d{\bf u}
\nonumber
\\
&\approx
\displaystyle
\sum_{i=1}^n\left[\log\left(g\left(\lambda {\bf A}^{-1} \left( {\bf x}_i - {\bm \mu} \right)\right)\right) +
\log\left(El \left( {\bf x}_i;  {\bm \mu}, {\bf A} {\bf A}^T, \nu \right)\right) \right]
\nonumber
\\
&
\displaystyle
\qquad
-n \log \sum_{j=1}^m g_m\left(\lambda {\bf u}_j\right) - n \log m,
\label{eq:QMLE}
\end{align}
where ${\bf u}$ is a set of random variables of size $m$ following pdf $El \left( {\bf u};  {\bf 0}, {\bf I}_k, \nu\right)$.

With this deterministic conditional quasi-log-likelihood function, more sophisticated optimisation algorithms can be used.
In this paper, the BFGS algorithm, an efficient and readily available quasi-Newton method implemented
in the $optim$ method in R (R Core Team, 2019) was used.
For the shape parameter, a first order stochastic gradient method can be used,
because the shape parameter is usually in low dimension, and the log-likelihood function is convex
in many cases such as for $t$ and power exponential.
The proposed QMLE/SGD algorithm is as follows:
\begin{enumerate}

\item
set $t = 0$, and initial parameters ${\bm\mu}^{(0)}$, ${\bf A}^{(0)}$, $\lambda^{(0)}$, $\nu^{(0)}$;

\item
generate ${\bf u}^{(t)}$ of size $m$ with $El \left({\bf 0}, {\bf I}_k, \nu^{(t)}\right)$;

\item
set the quasi-likelihood function with ${\bf u}^{(t)}$ according to (\ref{eq:QMLE});

\item
update parameters ${\bm \mu}^{(t+1)}$, ${\bf A}^{(t+1)}$, $\lambda^{(t+1)}$ by `loosely' maximising the conditional quasi-log-likelihood
given $\nu = \nu^{(t)}$ via the BFGS algorithm;

\item
obtain the conditional MLE, $\nu^{(t+1)}$, via SGD;

\item
set $t = t + 1$;

\item
repeat step 2 to 6 for a fixed number of iterations or until the convergence conditions are satisfied.

\end{enumerate}
Note that in step 4, the conditional quasi-log-likelihood is only `loosely' maximised.
It is because the quasi-log-likelihood function can be easily overfitted, and is generally unbounded.
For instance, consider the case ${\bf A} = {\bf I}$ and only one Monte carlo sample $u$ is used, the quasi-log-likelihood becomes
\begin{align*}
\displaystyle
\log QL \left(\lambda, \mu; u\right) = -\sum_{i=1}^n\frac {\left(x_i - \mu \right)^2}{2} +
\sum_{i=1}^n -\log\left\{ 1+\exp \left[-\lambda \left(x_i - \mu\right) \right] \right\} + n \log\left[1+\exp(-\lambda u)\right] + C.
\end{align*}
Unless $u = 0$, it can be shown that  $\log QL(\lambda, \mu;  u)$ is unbounded.
When $u$ is negative, we can set $\mu$ arbitrarily such that $x_i - \mu$ is positive for all $i \in 1,\ldots, n$,
and $\lambda$ is arbitrarily large.
As a result the QMLE can be approximated as
\begin{align*}
\displaystyle
\log QL \left( \lambda, \mu; u \right) \approx -\sum_{i=1}^n\frac {\left(x_i - \mu \right)^2}{2} - n \lambda u + C.
\end{align*}
Therefore, $\log QL(\lambda, \mu;  u)$ $\rightarrow \infty$ as $\lambda \rightarrow \infty$.
Similarly, if $u$ is positive, $\log QL(\lambda,\mu)$ $\rightarrow \infty$ as $\lambda \rightarrow -\infty$ for some arbitrarily large $\mu$.

The Monte carlo sample size required to reduce this QMLE bias grows exponentially with the number of dimensions,
and it can cause significant computational burden in higher dimensional problems.
Hence, given a limited Monte carlo sample size, overfitting of QMLE can be avoided by only `loosely' maximising the quasi-log-likelihood.
In case of BFGS, a small number of iterations can be set,
forcing the optimisation to stop prematurely before it starts to overfit.
It should be noted that BFGS with one iteration becomes gradient descent,
i.e. it becomes a first order method with no curvature information incorporated.
To demonstrate the performance of the proposed QMLE method, GMST-Logistic as described in Section \ref{section:sgd}
is applied to the river flow data set described in Table \ref{tab:river}.

\begin{figure}[H]
\includegraphics[width=15cm]{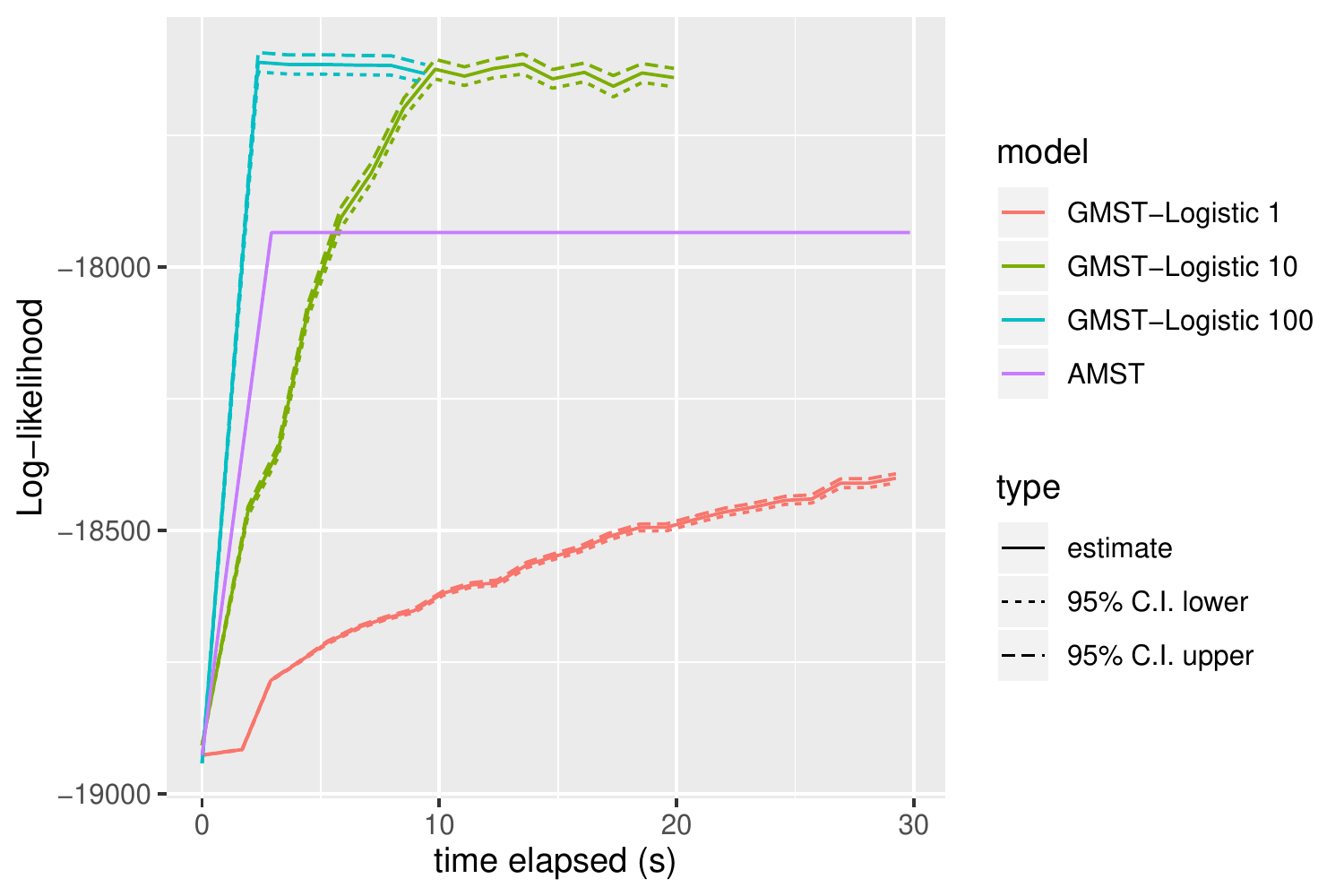}
\caption{Log-likelihood plot against computational time elapsed using QMLE/SGD method with 10,000 Monte carlo samples,
on the dataset containing the $\log$ river flows data recorded in the first two stations in Table \ref{tab:river}.
GMST-Logistic $p$ denotes the GMST-logistic fitted with QMLE/SGD algorithm using BFGS with maximum $p$ iterations.}
\label{fig:2d}
\end{figure}

\begin{figure}[H]
\includegraphics[width=15cm]{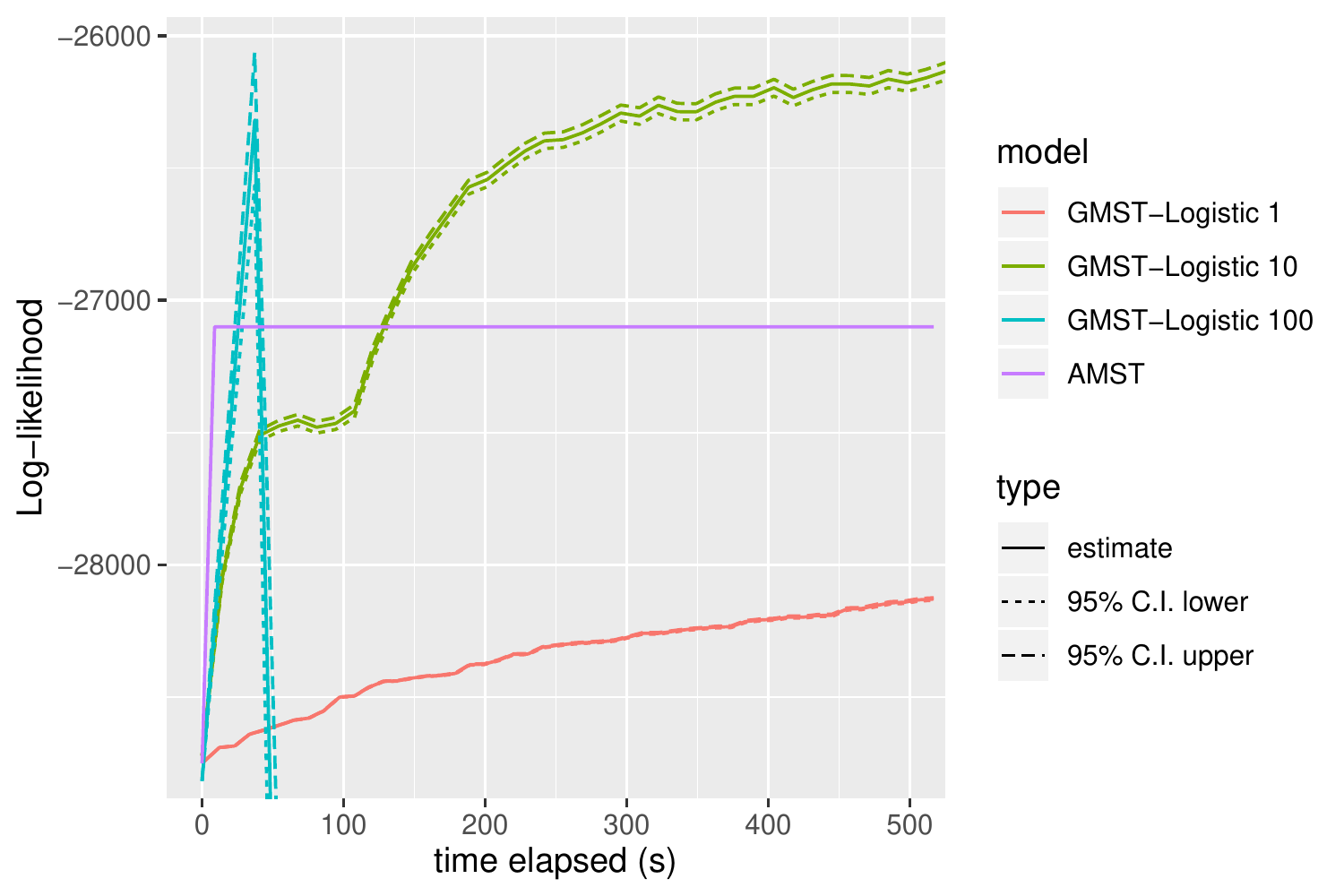}
\caption{Log-likelihood plot against computational time elapsed using QMLE/SGD method with 80,000 Monte carlo samples,
on the dataset containing the $\log$ river flows data recorded in the first five stations in Table \ref{tab:river}.
GMST-Logistic $p$ denotes the GMST-logistic fitted with QMLE/SGD algorithm using BFGS with maximum $p$ iterations.}
\label{fig:5d}
\end{figure}

GMST-Logistic using different maximum number of BFGS iterations was fitted to the first two dimensions and first five
dimensions of the river flow data set using Monte carlo sample sizes of 10,000 and 80,000, respectively.
Figure \ref{fig:2d}, and \ref{fig:5d} plot the log-likelihood achieved against time elapsed.
We can see from Figure \ref{fig:2d} that both GMST-Logistic with maximum BFGS iterations of 10 and 100 achieved
significantly higher log-likelihood than AMST within a reasonable time frame.
However, when a maximum BFGS iterations of one is used, the log-likelihood function increased significantly slower.
Nevertheless, the performance of the algorithm deteriorates significantly when 100 maximum BFGS iterations are used
for the five dimensional data set.
We can see from Figure \ref{fig:5d} that the log-likelihood fluctuates markedly when a maximum BFGS iterations of 100 was  used,
due to overfitting of QMLE.
When a maximum BFGS iterations of 10 was used instead, reasonable performance can be seen.

Optimising the number of BFGS iterations and Monte carlo sample sizes is a nuisance and time consuming task.
For moderately low dimensional ($d \le 10$) problems, we found that using BFGS iterations
of 10 and Monte carlo sample size of 10,000 would generally give reasonable performance for GMST-Logistic.
Performance of this setting on other variants of the SELIS has  been tested.
The conclusions were similar.

\section{Numerical examples}
\label{section:examples}

In this section, the ability of  the SELIS to approximate multivariate skew distributions is demonstrated.
Apart from the river flow data set described in Table \ref{tab:river},
we also consider the AIS data set (see Cook and Weisberg (1994) for description)
that was demonstrated in Azzalini and Capitanio (1999, 2003).
Due to the huge flexibility and varying specifications of  the SELIS, only GMST-Logistic with square skewing matrix is demonstrated in this paper.
For comparison, AMST and the special case of GMST-Logistic with diagonal skewing matrix (GMST-Logistic-D, hereafter) are considered.
MLE of both AMST and GMST-Logistic-D can be obtained rather quickly using $optim$ in R (R Core Team, 2019)
due to their closed-form pdfs.
More complex canonical skew distributions are not considered in this section,
since we were unable to find efficient methods or available packages in R that allow
efficient ML estimations for these distributions for the considered data sets.

\subsection{River flow data set}

In previous examples we have shown that GMST-Logistic gives significantly better
fits than AMST on lower dimensional subsets of this data set.
Here, the GMST-Logistic is fitted to the full ten dimensional data set using maximum BFGS iterations of 10 and 10000 Monte carlo samples.
Maximum log-likelihood, AIC and BIC obtained for each model on the raw data set and logarithm
of the data set are shown in Tables \ref{tab:fit_river} and \ref{tab:fit_logriver}, respectively.
In both cases, we see that both GMST-Logistic and GMST-Logistic-D give significantly better fits
than AMST in terms of AIC and BIC.
The time spent to fit GMST-Logistic-D in the second case is lower than AMST,
despite providing a huge surplus in log-likelihood.

A major reason for better fit is that GMST-Logistic and GMST-Logistic-D are more capable
to model data that has a weakly dependent skewing structure.
For instance, AMST can not adequately approximate independent skew $t$ distributions,
due to its univariate skewing function.
On the other hand, although GMST-Logistic can not model independence due to tail dependence of multivariate $t$,
having independent skewing functions allows it to approximate a weakly dependent skewing structure.
We loosely define a weakly dependent skewing structure as
\begin{align*}
\displaystyle
skewness(X | Y) \sim skewness(X).
\end{align*}
Clearly, AMST is unable to exhibit such a weakly dependent skewing structure.
Its conditional pdf of $X | Y$ is
\begin{align*}
\displaystyle
f_{X}({\bf x}|Y = {\bf y}) \propto T_1({\bf x} + {\bf B} {\bf y})t({\bf x} | Y = {\bf y})
\end{align*}
for some skewing matrix ${\bf B}$ of appropriate size.
While for GMST-Logistic (and the SELIS in general), the conditional pdf of $X|Y$ with appropriate skewing matrix and scale matrix can be written as
\begin{align*}
\displaystyle
f_{X}(x|Y = y) \propto g(x)t(x | Y = y).
\end{align*}
This allows  the SELIS to better approximate a weakly dependent skewing structure.

\begin{singlespace}
\begin{table}[H]
\centering
\begin{tabular}{@{}ccccc@{}}
\toprule
Model           & Log-likelihood & AIC    & BIC    & runtime(s)
\\
\midrule
AMST            & -84152         & 168456 & 168979 & 7.94
\\
GMST-Logistic-D & -63733         & 127617 & 128140 & 16.41
\\
GMST-Logistic   & -63545         & 127333 & 128165 & 52.50
\\
\bottomrule
\end{tabular}
\caption{Fitted log-likelihood, AIC, BIC and runtime cost by different models for the raw river flow data set.
Log-likelihood, AIC, and BIC are rounded to the nearest digit, runtime is rounded to 2 decimal places.}
\label{tab:fit_river}
\end{table}
\end{singlespace}

\begin{singlespace}
\begin{table}[H]
\centering
\begin{tabular}{@{}ccccc@{}}
\toprule
Model           & Log-likelihood & AIC   & BIC   & runtime(s)
\\
\midrule
AMST            & -48127         & 96406 & 96928 & 8.30
\\
GMST-Logistic-D & -47488         & 95128 & 95650 & 7.11
\\
GMST-Logistic   & -46720         & 93683 & 94515 & 129.46
\\
\bottomrule
\end{tabular}
\caption{Fitted log-likelihood, AIC, BIC and runtime cost by different models for the logarithm of the river flow data set.
Log-likelihood, AIC, and BIC are rounded to the nearest digit, runtime is rounded to 2 decimal places.}
\label{tab:fit_logriver}
\end{table}
\end{singlespace}

\begin{figure}[H]
\centering
\includegraphics[width=0.49\linewidth]{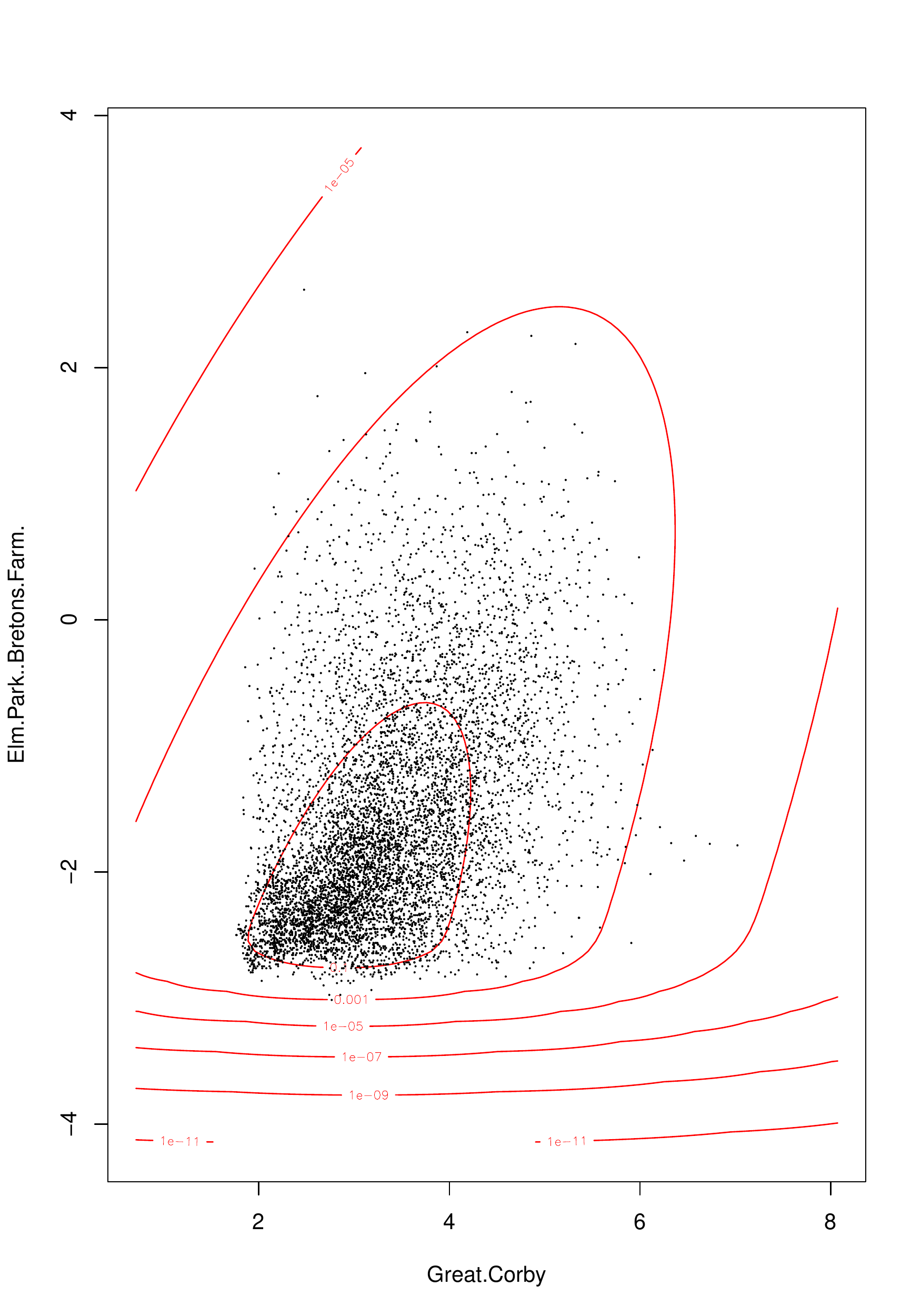}
\includegraphics[width=0.49\linewidth]{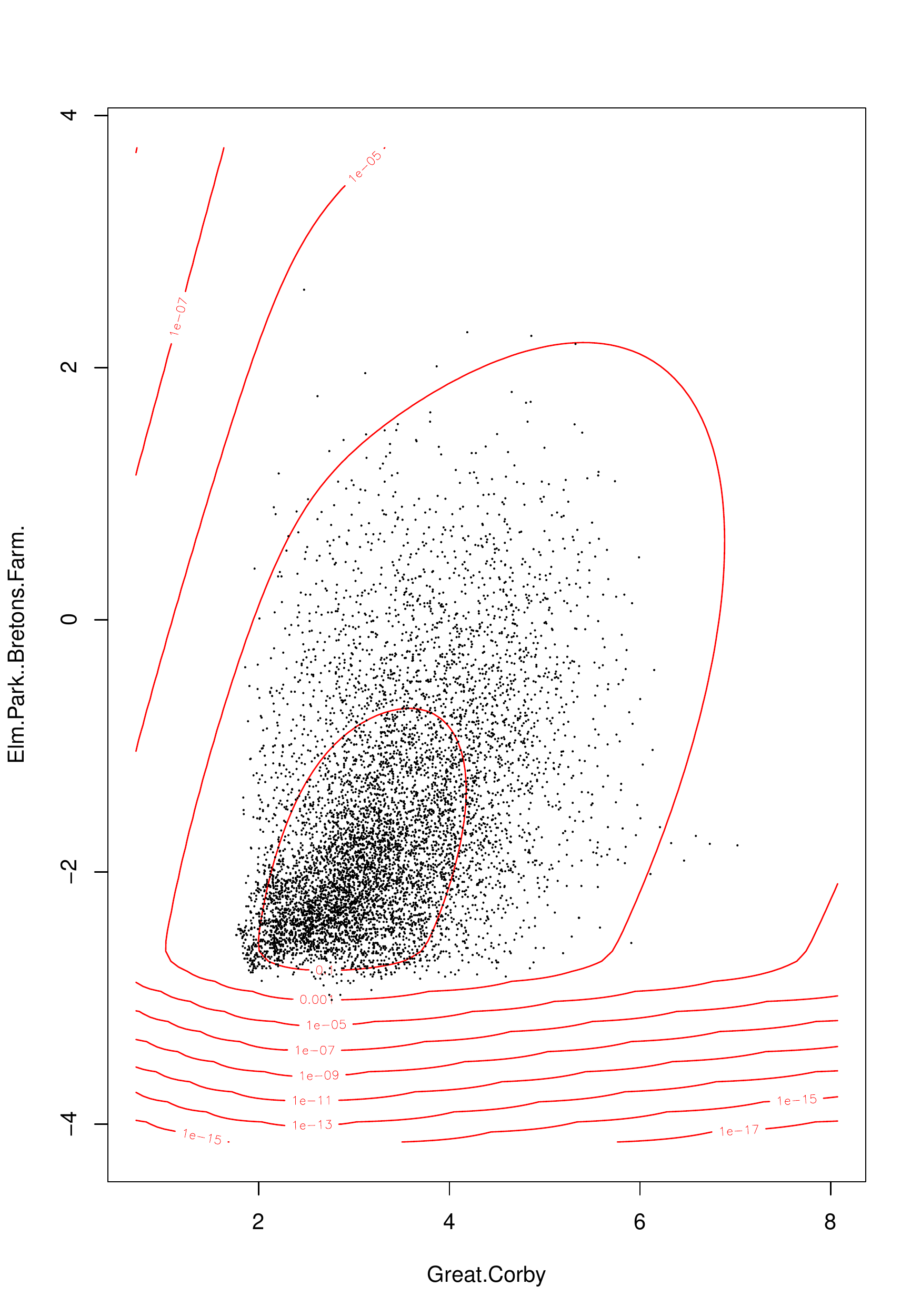}
\caption{Density contour plots of the fitted multivariate skew $t$-distributions on $\log$ daily river flow of rivers 9 and 10.
Left : AMST; Right : GMST-Logistic.}
\label{fig:rivercompare}
\end{figure}

For the river flow data set, although all data are obtained from rivers in the UK,
some stations are far apart and are affected by drastically different factors despite showing strong skewnesses.
Weaker dependent skewing structures can be expected from rivers that are further apart.
To illustrate the ability of GMST-Logistic and AMST to model this property,
both distributions were fitted to the 9\textsuperscript{th} and 10\textsuperscript{th} rivers,
recorded in the Great Corby station and Elm Park (Bretons Farm) station, respectively.
Figure \ref{fig:rivercompare} plots the density contour plots of the fitted AMST and GMST-Logistic.
It can be observed that GMST-Logistic gives graphically a much better fit than AMST,
as we can see there does not appear to be strong skewness dependence in the data set,
i.e. there appears to be some constant soft lower bounds for both variables.

\subsection{AIS data set}

The AIS data set contains eleven biomedical variables of 202 athletes collected at the Australian Institute of Sports.
The eleven variables are red blood cell count, white blood cell count, hematocrit, hemoglobin, plasma ferritin concentration,
body mass index, sum of skin folds, body fat percentage, lean body mass, height and weight.
Maximum log-likelihood, AIC and BIC obtained for each model on the raw data set and logarithm of the data set
are shown in Table \ref{tab:fit_ais} and \ref{tab:fit_logais}, respectively.

\begin{singlespace}
\begin{table}[H]
\centering
\begin{tabular}{@{}ccccc@{}}
\toprule
Model           & Log-likelihood & AIC   & BIC   & runtime(s)
\\
\midrule
AMST            & -4964          & 10106 & 10401 & 0.38
\\
GMST-Logistic-D & -4856          & 9890  & 10184 & 0.48
\\
GMST-Logistic   & -4856          & 10000 & 10477 & 33.15
\\
\bottomrule
\end{tabular}
\caption{Fitted log-likelihood, AIC, BIC and runtime cost by different models for the AIS data set.
Log-likelihood, AIC, and BIC are rounded to the nearest digit, runtime is rounded to 2 decimal places.}
\label{tab:fit_ais}
\end{table}
\end{singlespace}

\begin{singlespace}
\begin{table}[H]
\centering
\begin{tabular}{@{}ccccc@{}}
\toprule
Model           & Log-likelihood & AIC   & BIC   & runtime(s)
\\
\midrule
AMST            & 3484           & -6790 & -6496 & 0.18
\\
GMST-Logistic-D & 3558           & -6937 & -6643 & 0.34
\\
GMST-Logistic   & 3559           & -6831 & -6355 & 34.90
\\
\bottomrule
\end{tabular}
\caption{Fitted log-likelihood, AIC, BIC and runtime cost by different models for the logarithm of the AIS data set.
Log-likelihood, AIC, and BIC are rounded to the nearest digit, runtime is rounded to 2 decimal places.}
\label{tab:fit_logais}
\end{table}
\end{singlespace}

In this case, both GMST-Logistic-D and GMST-Logistic provide better fits than AMST in term of log-likelihood.
Both AIC and BIC, however, are worse in case of GMST-Logistic, and GMST-Logistic only improved log-likelihood by 1 from GMST-Logistic-D.
We believe that this is due to the small sample size, for a eleven dimensional data set, 202 data points are inadequate
to exhibit a detailed relationship between variables.
Overall, there is very strong evidence that GMST-Logistic-D gives a significantly better fit for this data set.

\section{Concluding remarks}
\label{section:conclusion}

In conclusion, we have introduced a class of distributions, SELIS, that is capable to approximate more complex skewing structures.
An efficient estimation method has been proposed and demonstrated.
We have shown that not only the proposed class is very flexible,
it can be fitted within a much desirable time frame and suffer less from the curse of dimensionality
compared to other skew distributions of similar complexity.
In addition, we have shown that the nested form of  the SELIS with diagonal skewing matrix can
often provide much better fits than other skew distributions of similar complexity.
Even better, our study shows that it can be fitted as quickly as AMST,
which is arguably the simplest form of multivariate skew $t$ distributions.
We believe, and our study strongly suggests that  the SELIS and its diagonal nested form
have huge potential in other applications involving statistical analyses.
For example, it can be applied in finite mixture models to help conduct better cluster analysis or density estimation.
However, there are a few clear limitations of  the SELIS.
For instance, marginal distributions of  the SELIS are generally not analytical,
hence it is very hard to visualise the fitted distributions in lower dimensional contour plots or density plots.
Also, the absence of convolution type stochastic representation increases the difficulties to study its statistical properties.

The study of  the SELIS is far from complete, there are many aspects of  the SELIS that can be studied further.
For instance, due to the non-analytical pdf, statistical properties of  the SELIS have not been studied in detail.
Moreover, there are many potential future directions that can improve the estimation process,
e.g. implementation quasi-Newton method and efficient approximation of the normalising integral.

\end{document}